\documentclass[%
 reprint,
superscriptaddress,
 amsmath,amssymb,
 aps,
]{revtex4}
\usepackage{hyperref}
\hypersetup{colorlinks=true,citecolor=red,linkcolor=blue,urlcolor=blue}
\usepackage[dvipsnames]{xcolor}
\usepackage{amsmath}

\usepackage{xcolor}
\usepackage{enumitem}
\usepackage{float}
\usepackage{graphicx}
\usepackage{dcolumn}
\usepackage{bm}

\usepackage{orcidlink}
\begin{document}

\preprint{APS/123-QED}

\title{Exploring Wormhole Structures within the Framework of $f(R,L_{m})$ Gravity}
\author{A. S. Agrawal\orcidlink{0000-0003-4976-8769}}
\email{asagrawal.sbas@jspmuni.ac.in}
\affiliation{Department of Mathematics, School of Computational Science, Faculty of Science and Technology, JSPM University Pune-412207, India.}
\author{Baljeet Kaur Lotte\orcidlink{0000-0001-5226-9640}}
\email{ballotte21@gmail.com}
\affiliation{Department of Physics, Odisha University of Technology and Research,Bhubaneswar 751029, India}
\author{Reshma Kapse\orcidlink{0009-0003-9389-0670}}
\email{reshmakapse10@gmail.com}
\affiliation{Department of Mathematics, School of Computational Science, Faculty of Science and Technology, JSPM University Pune-412207, India.}
\author{B. Mishra\orcidlink{0000-0001-5527-3565}}
 \email{bivu@hyderabad.bits-pilani.ac.in}
 \affiliation{Department of Mathematics,
Birla Institute of Technology and Science-Pilani, Hyderabad Campus, Jawahar Nagar, Kapra Mandal, Medchal District, Telangana 500078, India.}

\date{\today}

\begin{abstract}
In this work, we investigate wormhole geometries within the framework of $f(R,\mathcal{L}_{m})$ gravity by considering a specific form of the model. From the corresponding field equations, the shape function is derived, and the traversability conditions are examined for suitable choices of the model parameters. The obtained shape function is shown to satisfy all the necessary requirements for a traversable wormhole, including the energy conditions. The geometric properties of the wormhole are analyzed in detail, and particular attention is given to the requirement that the proper radial distance $l(r)$ remains finite throughout the space-time, thereby ensuring the consistency of the geometry.
\end{abstract}
\maketitle

\section{Introduction} \label{Sec:I}
General Relativity (GR) and Quantum  Field Theory (QFT) form the basis of our understanding on elementary particles \cite{Visser:1995cc}. The framework of these theories has provided accurate and consistent descriptions of nature within its respective domain of applicability. On macroscopic and cosmological scales, GR has successfully passed every experimental and observational test of gravity. Whereas, the quantum field theory, realized in the standard model of particle physics, offers successful account of subatomic phenomena. Both frameworks stand as monumental milestones to uncover the fundamental laws of the Universe. Following GR, Flamm \cite{flamm1916} was among the first to examine a potential solution to Einstein’s field equations and suggested a wormhole like structure. However, his solution was later found to be unstable. The notable finding in this direction was the Einstein–Rosen bridge \cite{Einstein73} and further Wheeler \cite{Wheeler1955, Wheeler:1957mu} proposed wormhole and the concept of spacetime foam. Morris and Thorne \cite{Morris95} weigh on the idea of whether wormholes can be traversed or if they exist.\\

According to GR, matter curves spacetime and sufficiently massive objects can form singularities where spacetime breaks down, leading to black holes. If singularities are avoided, spacetime may allow for traversable wormhole tube-like structures with flat regions at both ends, connected by a throat that may be static or evolving. For stability within GR, wormholes generally require exotic matter that violates the null energy condition (NEC) \cite{Visser1995, Hochberg21}, unlike ordinary matter, which satisfies it.  A wormhole geometry can be visualized by imagining a sheet of paper representing space. If the paper is folded and a hole is pierced through it, two distant points on the sheet become directly connected. Similarly, in spacetime, a wormhole acts as a shortcut linking remote regions. In GR, traversable wormholes typically require exotic matter that violates the classical energy conditions. To overcome this difficulty, several modified gravity theories have been investigated, such as $f(R)$, $f(R,T)$, $f(R,\mathcal{L}_m)$, and $f(R,\mathcal{L}_m,T)$, with the aim of supporting wormhole geometries with little or no exotic matter ~\cite{De_Felice_2010, sotiriou2010f, Harko2011, HarkoLobo2010, haghani2021generalizing}. \\ 

One of the earliest extensions of GR is $f(R)$ gravity, in which the action depends on an arbitrary function of the Ricci scalar $R$. This modification introduces higher-order corrections and has been employed to describe cosmic acceleration as well as to construct wormhole solutions~\cite{Capozziello2002CurvatureQuintessence, NojiriOdintsov2011UnifiedHistory}. Lobo and Oliveira~\cite{Lobo_2009} showed that in certain $f(R)$ wormhole configurations, ordinary matter can respect the energy conditions, while the required violation is provided by higher-order curvature terms, interpreted as an effective gravitational fluid. Pavlovic and Sossich~\cite{Pavlovic_2015} further demonstrated that traversable wormholes may exist without the need for exotic matter, while examining weak energy condition bounds and curvature effects near the throat. Agrawal {\textit et al.}~\cite{Agrawal22} reported analytic wormhole solutions in $f(R)$ gravity using a class-I approach, checked null and weak energy conditions, constrained parameters, compared shape functions with the Gronwall–Bellman bound, and studied tidal forces for traversability. Mishra {\textit et al.}~\cite{Mishra21} developed explicit wormhole models in $f(R)$ gravity by considering particular functional forms of $f(R)$ linked to model parameters and explored the physical viability of several shape functions. Together, these works provide a broad set of tools for constructing wormholes in $f(R)$ gravity. Another modified theory of interest is $f(R,T)$ gravity, where the action depends on the Ricci scalar $R$ and the trace of the energy–momentum tensor $T$~\cite{Harko2011}. The matter–geometry coupling introduces additional degrees of freedom, allowing ordinary matter to satisfy energy conditions while the effective sector mimics exotic contributions, leading to studies on stability, energy conditions, and traversable wormhole solutions~\cite{Zubair2016, Mishra_2020}. In $f(R,\mathcal{L}_m)$ gravity, the action includes a non-minimal curvature–matter coupling, and the resulting field equations imply that test particles follow non-geodesic motion due to an extra force~\cite{Harko:2010:70}. This coupling naturally allows the violation of the null energy condition and the realization of wormhole geometries. Garcia and Lobo~\cite{Garcia_2010} demonstrated that the effective stress–energy tensor in this theory can support wormholes without exotic matter. Avelino {\textit et al.}~\cite{Avelino_2018} provided further insight by showing that, for fluids, the average matter Lagrangian coincides with the trace of the energy–momentum tensor, recovering $f(R)$ as a special case. More recently, Shukla {\textit et al.}~\cite{Shukla_2025} studied a flat FRW universe in this framework, using phase-plane analysis to show a transition from early deceleration to present cosmic acceleration.  

The most general extension studied so far is $f(R,\mathcal{L}_m,T)$ gravity, where the action depends simultaneously on $R$, $\mathcal{L}_m$, and $T$. This stronger coupling has been shown to provide stable and traversable wormhole solutions while reducing violations of energy conditions~\cite{haghani2021generalizing}. Bhattacharjee {\textit et al.}~\cite{Bhattacharjee:2019}, Zubair and Azmat~\cite{Zubair:2017}, and Moraes {\textit et al.}~\cite{Moraes:2024} have reported explicit wormhole models in this setting. In particular, Moraes {\textit et al.} demonstrated that a simple $f(R,\mathcal{L}_m,T)=R+\alpha \mathcal{L}_m + \beta T$ model yields solutions that fulfill traversability criteria without exotic matter. In this paper, we shall examine the possibility to construct wormhole solutions without requiring exotic matter in $f(R,\mathcal{L}_m)$  gravity and to analyze the throat structure of the wormhole, and test the validity of the energy conditions. In Sec.\ref{Sec:II}, we present the basic formalism of the field equations for wormhole geometry. In Sec. \ref{Sec:III}, we develop the shape function together with the conditions imposed on the shape function and the corresponding energy conditions. In Sec.\ref{Sec:IV}, the wormhole geometry is constructed using the shape function obtained in Sec. \ref{Sec:II}. In Sec.\ref{Sec:V}, we provide the results and discussion.

\section{The $f(R,\mathcal{L}_m)$ gravity}\label{Sec:II}

The action of $f(R,\mathcal{L}_m)$ gravity \cite{Harko:2010:70} given as,

\begin{equation}\label{cfrl1}
S=\int d^4x \sqrt{-g}f(R,\mathcal{L}_m),
\end{equation}
where the Ricci scalar $R$ can be obtained by contraction of the Ricci tensor $R_{\mu \nu}$. Primarily, the curvature of spacetime is  characterized by the Ricci tensor that describes how the volume in spacetime deviates from the flatness. The metric tensor $g_{\mu \nu}$ determines the geometric structure of spacetime by defining distances and angles and its determinant $g=|{g_{\mu \nu}|}$ has significant contribution in the formulation of volume element of spacetime. Using variational principle in action \eqref{cfrl1}, the field equations for $f(R,\mathcal{L}_m)$ gravity \cite{Harko:2010:70} can be obtained as,
\begin{eqnarray}\label{Eq:field1}
    R_{\mu\nu}f_R+[g_{\mu\nu}\nabla_\mu\nabla^\mu-\nabla_\mu\nabla_\nu]f_R 
    -\frac{1}{2}f(R,\mathcal{L}_m)g_{\mu\nu}=\frac{1}{2}f_{\mathcal{L}_m}[T_{\mu\nu}-\mathcal{L}_mg_{\mu\nu}]\,.\label{cfrl2}
\end{eqnarray}
The function $f(R,\mathcal{L}_{m})$ is expressed as a combination of two arbitrary functions, $f_{1}(R)$ and $f_{2}(\mathcal{L}_{m})$. Here, $f(R)$ depends on $R$, which characterizes spacetime curvature, while $f_{2}(\mathcal{L}_{m})$ is a function of the matter Lagrangian density $\mathcal{L}_{m}$, describing the distribution of matter. For brevity, we denote 

\begin{eqnarray*}
    f_{R} = \frac{\partial f(R,\mathcal{L}_m)}{\partial R}, 
\qquad 
f_{\mathcal{L}_m} = \frac{\partial f(R,\mathcal{L}_m)}{\partial \mathcal{L}_m},
\end{eqnarray*}
where $f_{R}$ and $f_{\mathcal{L}_m}$ denotes respectively the rate of change of $f(R,\mathcal{L}_m)$ with respect to $R$ and $\mathcal{L}_m$. Now, the field equation \eqref{Eq:field1} can be rewritten as,
\begin{eqnarray}\label{Eq:field2}
f_{R}R+3\Box f_{R}-2f=f_{\mathcal{L}_{m}}\left(\frac{1}{2}T-2\mathcal{L}_{m}\right).    
\end{eqnarray}
By operating the d'Alembert operator $\Box$ on Eq.~\eqref{Eq:field1} and Eq.~\eqref{Eq:field2}, we obtain
\begin{eqnarray}
f_{R}\left[R_{\mu \nu}-\frac{R}{3} g_{\mu \nu}\right]+\frac{f}{6}-\nabla_{\mu}\nabla_{\nu}f_{R}
=\frac{f_{\mathcal{L}_{m}}}{2}\left[T_{\mu \nu}-\frac{1}{3}(T-\mathcal{L}_{m}) g_{\mu \nu} \right]\,. \label{Eq:field3}  \end{eqnarray}

The covariant divergence of the field equation~\eqref{Eq:field1}, in conjunction with the relevant identity, yields the divergence of the energy--momentum tensor $T_{\mu \nu}$~\cite{Harko_2020} as,
\begin{eqnarray}
\nabla^{\mu}T_{\mu \nu}&=& 2\nabla^{\mu}\ln[f_{\mathcal{L}_{m}}]\frac{\partial \mathcal{L}_{m}}{\partial g^{\mu \nu}}.
\end{eqnarray}
The energy-momentum conservation of the matter field, expressed as $\nabla^{\mu}T_{\mu \nu}=0$, leads to a functional relationship between the Lagrangian density and the function $f_{\mathcal{L}_{m}}$

\begin{equation}\label{cfrl3}
	\nabla^\mu\ln f_{\mathcal{L}_m}(R,\mathcal{L}_{m})=0\,.
\end{equation}
Once the matter Lagrangian density is determined, an appropriate specification of the function $f(R,\mathcal{L}_m)$ can, in principle, be employed to construct conservative frameworks incorporating arbitrary curvature-matter interactions.

The Morris--Thorne metric~\cite{Morris95} serves as the basis for traversable wormholes and is defined by the conditions on the redshift and shape functions which ensure traversibility and stability. The Morris--Thorne metric is,
\begin{equation}\label{MTM}
    ds^2=-e^{2\Phi(r)}dt^2+\bigg[1-\frac{b(r)}{r}\bigg]^{-1}dr^2+r^2(d\theta^2+\sin^2\theta d\phi^2),
\end{equation}

where $(t,r,\theta,\Phi)$ represents the usual space–time spherical coordinates. $\Phi(r)$ and $b(r)$ are arbitrary functions of the radial coordinate $r$ and known as redshift function and shape function respectively. The redshift function determines the gravitational redshift whereas the shape function determines the spatial shape of the wormhole. The following conditions must be satisfy,
\begin{itemize}
    \item The redshift function $\Phi(r)$ must be finite everywhere for $r \geq r_{0}$ in order to avoid an event horizon, so that
    \begin{equation}\label{RF_I}
                e^{\Phi(r)} > 0 \quad \text{for all } r > r_{0}.
    \end{equation}
    Moreover,
    \[
        \lim_{r \to +\infty} \Phi(r) = \Phi_{0},
    \]
    where $\Phi_{0}$ is finite and real.
    
    \item The flare-out condition must hold,
    \begin{equation}
        \frac{b(r) - b'(r)r}{2b^{2}(r)} > 0 \quad \text{at throat} \quad r_{0} \quad \text{or near the throat } r = r_{0}.
    \end{equation}
    
    \item The above condition implies that for all $r \geq r_{0}$:
    \begin{equation}
        b(r_{0}) = r_{0}, \qquad b'(r_{0}) < 1.
    \end{equation}
    Furthermore, for $r > r_{0}$ we have
    
    \begin{equation}
        b(r) < r\,, \qquad  b'<\frac{b(r)}{r}\,.\label{SF_III}
    \end{equation}
The prime denotes differentiation of $b(r)$ with respect to the radial coordinate $r$, and it must satisfy the geometric condition $b'(r) < \frac{b(r)}{r}$, at all points for a traversable wormhole. This ensures that the shape function maintains the appropriate curvature required for wormhole stability and traversability. Furthermore, the condition $b(r) < r$, must hold at all points away from the throat $r_{0}$.
    \item To ensure the asymptotic behavior, the shape function $b(r)$ must fulfill
    \begin{equation}
        \lim_{r \to +\infty} b(r) = \text{finite}.
    \end{equation}
To establish stability and traversability, the wormhole must satisfy specific conditions at the throat $r_{0}$. These include the flare-out condition, ensuring that the geometry spreads outward from the throat, and the requirement that the shape function fulfills $b(r_{0}) = r_{0}$.
\end{itemize}
Now, the matter content of the wormhole described by the energy–momentum tensor can be expressed as an anisotropic distribution of matter as,
\begin{equation}\label{EMT_I}
T_{\mu \nu}=(\rho +p_{t})u_{\mu}u_{\nu}+p_{t}g_{\mu \nu}+(p_{r}-p_{t})x_{\mu}x_{\nu}\,,    
\end{equation}
where $x^{\mu}$ is a space-like unit vector satisfying $x^{\mu}x_{\mu} = 1$, $u^{\mu}$ is a time-like normalized vector representing the four-velocity of the fluid, 
with $u^{\mu}u_{\mu} = -1$ and orthogonal to $x^{\mu}$. 
$\rho$ denotes the energy density, and $p_{r}$ and $p_{t}$ are the radial and transverse pressures, respectively \cite{Bahamonde2016}. The trace of the energy--momentum tensor that provides a key relationship between the energy density and pressure components can be expressed as,
\begin{equation}\label{EMT_II}
    T^\mu_\nu=[-\rho,p_r,p_t,p_t]. 
\end{equation}
The field equations of $f(R,L_m)$ gravity for the Morris–Thorne wormhole metric can be derived and expressed as the energy density, radial pressure and traversal pressure respectively as,
\begin{subequations}
\begin{eqnarray}
\rho&=& \frac{f}{f_{\mathcal{L}_{m}}}-\frac{f_{R}}{r^{2}f_{\mathcal{L}_{m}}}\bigg[\big(r(b'-4)+3b\big)\Phi'-2r(r-b)\Phi'^{2} +2r(b-r)\Phi''\bigg]-\frac{f_{RR}}{r^{2}f_{\mathcal{L}_{m}}}\bigg[2r(r-b)R''-\nonumber \\
&&\big(r(b'-4)+3b\big)R'\bigg] -\frac{2f_{RRR}R'^{2}}{f_{\mathcal{L}_{m}}}\left(1-\frac{b}{r}\right)-\mathcal{L}_{m}\,,\nonumber \\
\\
p_{r} &=& - \frac{f}{f_{\mathcal{L}_{m}}} +\frac{bf_{R}}{r^{3}f_{\mathcal{L}_{m}}}\left(2r^{2}\Phi'^{2}+2r^{2}\Phi''-r\Phi'-2\right)+\frac{f_{R}}{r^{2}f_{\mathcal{L}_{m}}}\bigg(b'(r\Phi'+2)-2r^{2}(\Phi'^{2}+\Phi'')\bigg)\nonumber \\
&&+\frac{2f_{RR}R'}{f_{\mathcal{L}_{m}}}\left(1-\frac{b}{r}\right)\left(\Phi'+\frac{2}{r}\right)+\mathcal{L}_{m}\,, \\
p_{t}&=& -\frac{f}{f_{\mathcal{L}_{m}}}+\frac{f_{R}}{r^{3}f_{\mathcal{L}_{m}}}\bigg(b(2r\Phi'+1)+(b'-2r\Phi')r\bigg)+\frac{2f_{RR}}{r^{2}f_{\mathcal{L}_{m}}}\bigg[\frac{R'}{2}\big(r(2r\Phi'-b'+2)-b(2r\Phi'+1)\big)\nonumber \\
&&+r(r-b)R''\bigg] +\frac{2R'^{2}f_{RRR}}{f_{\mathcal{L}_{m}}}\left(1-\frac{b}{r}\right)+\mathcal{L}_{m}\,.
\end{eqnarray}
\end{subequations}
The Ricci scalar for a tideless wormhole, where $\Phi$ is constant, is given by $R = \frac{2b'(r)}{r^{2}}$. Accordingly, the field equation can be rewritten as
\begin{subequations}
\begin{eqnarray}
\rho&=& \frac{f}{f_{\mathcal{L}_{m}}}-\frac{f_{RR}}{r^{2}f_{\mathcal{L}_{m}}}\bigg[(r-b)\frac{24b'+4r(b^{(3)}r-4b'')}{r^3}-\big(r(b'-4)+3b\big)\frac{2 \left(r b''-2 b'\right)}{r^3}\bigg]\nonumber \\
&&-\frac{8f_{RRR}}{f_{\mathcal{L}_{m}}}\frac{(r b''-2 b')^{2}}{r^6}\left(1-\frac{b}{r}\right)-\mathcal{L}_{m}\,, \label{rho_a} \\
p_{r} &=& - \frac{f}{f_{\mathcal{L}_{m}}} -\frac{2bf_{R}}{r^{3}f_{\mathcal{L}_{m}}}+\frac{2b'f_{R}}{r^{2}f_{\mathcal{L}_{m}}}+\frac{8f_{RR}}{f_{\mathcal{L}_{m}}}\frac{ \left(r b''-2 b'\right)}{r^4}\left(1-\frac{b}{r}\right) +\mathcal{L}_{m}\,,\label{pr_a} \\
p_{t}&=& -\frac{f}{f_{\mathcal{L}_{m}}}+\frac{f_{R}}{r^{3}f_{\mathcal{L}_{m}}}(b+rb')+\frac{2f_{RR}}{r^{2}f_{\mathcal{L}_{m}}}\bigg[\frac{(r b''-2 b')}{r^3}\big(r(2-b')-b\big)+(r-b)\frac{12b'+2r(b^{(3)}r-4b'')}{r^3}\bigg]\nonumber \\
&& +\frac{8f_{RRR}}{f_{\mathcal{L}_{m}}}\frac{(r b''-2 b')^{2}}{r^6}\left(1-\frac{b}{r}\right)+\mathcal{L}_{m}\,, \label{pt_a}
\end{eqnarray}   
\end{subequations}

$b'$, $b''$ and $b^{(3)}$ respectively denotes $\frac{db}{dr}$, $\frac{d^{2}b}{dr^{2}}$ and $\frac{d^{3}b}{dr^{3}}$. To proceed further in the wormhole solution, we need some functional form of $f(R,\mathcal{L}_{m})$, which we shall discuss in the next section.

\section{Solutions to Wormhole in $f(R,\mathcal{L}_{m})$ Gravity}\label{Sec:III}
We consider the non-linear functional form of $f(R,\mathcal{L}_{m})$ as
\begin{equation}
f(R,\mathcal{L}_{m})=f_{1}(R)+[1+\lambda f_{2}(R)]\mathcal{L}_{m}, \label{frlm}
\end{equation}
 where $\lambda$ is the model parameter, and for $f_{1}(R)=\tfrac{R}{2}$ with vanishing $\mathcal{L}_{m}$, the model reduces to the standard equations of GR. However, the additional contribution is typically introduced to address the late-time accelerated expansion of the Universe. Here, we fix the functions $f_{1}(R)$ and $f_{2}(R)$ such that the resulting field equations yield improved phenomenological outcomes. 
Specifically, we set $f_{1}(R) = \tfrac{R^{2}}{2}$ and $f_{2}(R) = R$, where $\lambda$ denotes the coupling constant. Using this specific form of $f(R,\mathcal{L}_m)$, Eq. \eqref{rho_a} reduces to,

\begin{eqnarray}
\rho&=& \frac{R^{2}}{2(1+\lambda R)}-\frac{1}{r^{2}(1+\lambda R)}\bigg[(r-b)\frac{24b'+4r(b^{(3)}r-4b'')}{r^3}-\big(r(b'-4)+3b\big)\frac{2 \left(r b''-2 b'\right)}{r^3}\bigg]\,, \label{rho_a1} 
\end{eqnarray}   
Solving the above equation, we will get 
\begin{equation}
    \rho= \left[\frac{1}{r^{2}(r^{2}+2\lambda b')}\right]\left[2b'^{2}-\frac{1}{r}\left[(r-b)(24b'+4r\left(rb'''-4b'' \right))-2\left(r(b'-4)+3b\right)\left(rb''-2b'\right)\right]\right] \label{rhoa1}
\end{equation}
To solve Eq. \eqref{rhoa1}, we are intending to choose a shape function $b(r)$ that is quadratic, so that  $b'''(r) = 0$ and hence 
\begin{equation}
    b(r)=r_{0} -\frac{1}{2} (r-r_{0} ) (r_{0} +r-2 \alpha ) \label{b(r)_1},
\end{equation}
by choosing suitable values of $r_{0}$, $\alpha$, and $\lambda$, the energy density can be made to approach ${1}/{r^{2}}$. Since the term ${1}/{r^{2}}$ as a value of energy density always has a positive value, our goal is to select parameters such that the resulting shape function satisfies the traversability conditions of the wormhole, including the key requirement $b(r_{0}) = r_{0}$. One possible set of parameter values that meets all these conditions is
\begin{equation*}
   \alpha = 2 \quad \lambda = 6 \quad \text{and} \quad r_0 = 2. 
\end{equation*}
For the chosen parameters, the energy density, radial pressure, and tangential pressure can be defined as,
\begin{eqnarray}
    \rho &=& \frac{1}{r^{2}}\\
    p_{r} &=& \frac{2 r (r (r (2 r-31)+204)-480)+768}{r^4 ((r-12) r+24)}\\
    p_{t} &=& \frac{(29-2 r) r-56}{r^2 ((r-12) r+24)} 
\end{eqnarray}
Using Eq.~\eqref{b(r)_1} and the parameter set \(\alpha = 2 \) and \(r_0 = 2\), we now proceed to obtain a graphical representation of the shape function \(b(r)\), its derivative \(b'(r)\), and the ratio \(b(r)/r\). Additionally, we plot the reference line \(b(r) = r\) to visually verify the fulfillment of the condition \(b(r_0) = r_0\). These plots provide a qualitative confirmation that the chosen value of \(r_0\) is consistent with the fundamental requirement for a traversable wormhole throat. 

\begin{figure}[H]
    \centering
    \includegraphics[width=0.5\linewidth]{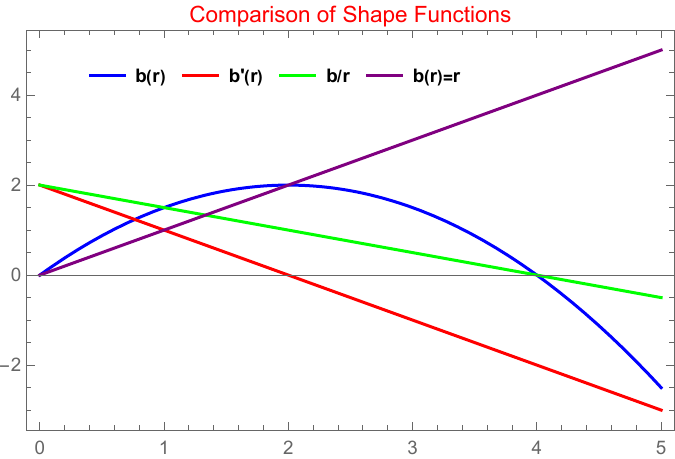}
    \caption{Graphical comparison of the shape function \(b(r)\), its first derivative \(b'(r)\), the ratio \(b(r)/r\), and the identity line \(b(r) = r\). }
    \label{fig:enter-label1}
\end{figure}
As the value \(r_0 = 2\) has been fixed, it is essential to visually verify whether this value satisfies the throat condition \(b(r_0) = r_0\). To this end, we plot the shape function \(b(r)\) alongside the reference line \(b(r) = r\)~[Fig.~\ref{fig:enter-label1}]. The point of intersection between these two curves serves as direct confirmation that the selected value of \(r_0\) fulfills the necessary condition for a traversable wormhole throat. The observed intersection at \(r_0 = 2\) confirms that \(b(r_0) = r_0\), thereby validating this fundamental requirement for the existence of a traversable wormhole. After substituting the above value for $b(r)$, we get the equation for energy density, radial pressure and the tangential pressure respectively as,

\begin{eqnarray}
\rho &=& \frac{6 \alpha r_0 (-2 \alpha + r_0 + 2) - 6 \alpha r^2 + r^3 + r \left(10 \alpha^2 + 2 \alpha (r_0 - 4) - r_0 (r_0 + 2)\right)}{r \left(2 \alpha \lambda + r^2 - 2 \lambda r\right)} ,\\
\nonumber\\
p_r &=& \frac{1}{r^5 \left(2 \alpha \lambda + r^2 - 2 \lambda r\right)} \Bigg[ 
\lambda \Big( 
r \big(32 (\alpha -1) \alpha^2 + 8 (5 \alpha -2) r^2 - 7 r^3 + 16 \alpha (3 - 4\alpha) r \big) + r_0^2 (3r - 4\alpha)^2 \nonumber \\
&& \qquad  - 2 (\alpha - 1) r_0 (3r - 4\alpha)^2 
\Big) + 2 r^2 \Big(
3 \alpha r_0 (-2\alpha + r_0 + 2) 
+ (4 - 7 \alpha) r^2 + 2 r^3  \nonumber \\
&&\qquad + r \big(7 \alpha^2 + 2 \alpha (r_0 - 4) - r_0 (r_0 + 2)\big) 
\Big) \Bigg], \\
\nonumber\\
p_t &=& \frac{1}{2 r^3 \left(2 \alpha \lambda + r^2 - 2 \lambda r\right)} \Bigg[
4 \alpha^2 (9 r_0 + 2) + 2 \alpha r_0 (\lambda - 9 (r_0 + 2)) - 16 \alpha - r_0 (r_0 + 2) \lambda 
+ r^2 (26 \alpha + 3 (\lambda - 4)) \nonumber \\
&& \qquad  - 4 r^3  + 4 r \left( -9 \alpha^2 - \alpha (2 r_0 + \lambda - 9) + r_0 (r_0 + 2) + 2 \right) 
\Bigg].
\end{eqnarray}

It has been observed that while the flare-out condition and the shape function are satisfied for the given parametric constant values, the energy conditions are not fulfilled. Therefore, in the following analysis, we aim to explore alternative sets of parametric values that can simultaneously satisfy the shape function requirements and the necessary energy conditions to achieve a viable traversable wormhole solution. As the shape function is independent of the parameter $\lambda$, we have the flexibility to choose $\lambda$ such that the energy conditions are satisfied. In the following, we examine the null energy condition in both radial and tangential directions, taking $\lambda$ in the range ($2.6$, $4$) and $r$ in the range ($2$,$5$).
  
\begin{figure}[H]
    \centering
    \includegraphics[width=0.45\linewidth]{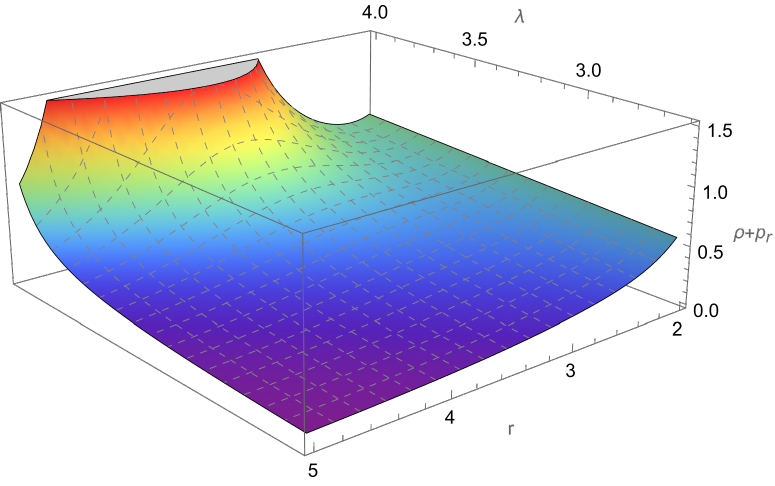}
    \includegraphics[width=0.45\linewidth]{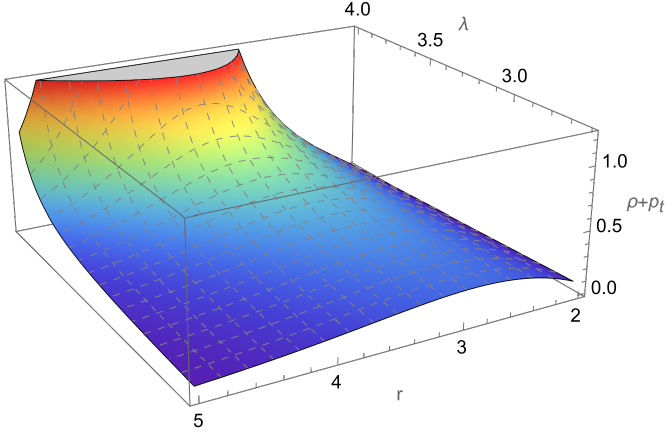}
    \caption{Graphical representation of $\rho + p_{r}$ (left) and $\rho + p_{t}$ (right) over the range $r \in (2,5)$ and $\lambda \in (2.6,4)$, illustrating that both quantities remain positive, thereby satisfying the Null Energy Condition (NEC).}
    \label{fig:enter-label2}
\end{figure}

In the above graphical representation~[Fig.\ref{fig:enter-label2}], we have adopted the parametric values $\{ r_{0}, \alpha \} = \{ 2, 2 \}$ and considered $\lambda \in [2.6, 4]$. For this configuration, it is observed that both null energy conditions, $\rho + p_{r}$ and $\rho + p_{t}$, are satisfied. However, even a slight increase or decrease in the value of $\lambda$ leads to the violation of one of these NECs. Consequently, we conclude that a stable, traversable wormhole solution in $f(R, \mathcal{L}_{m})$ gravity can exist without the necessity of exotic matter.
 
\section{ The Wormhole Structure} \label{Sec:IV}
In this section, we present the wormhole space-time through an embedding diagram and analyze the characteristics of the resulting shape function $b(r)$. Furthermore, we compute the proper radial distance and illustrate its extent.
Since we are working with spherically symmetric and static wormhole solutions, the solid angle element $d\Omega^2 \equiv d\theta^2 + (\sin\theta\, d\phi)^2$ can be written as
\begin{equation}
d\Omega^{2}= d\varphi^{2},    
\end{equation}
by restricting the analysis to equatorial plane $\theta = \pi/2$  given in the the generic line element \eqref{MTM}. Moreover, at a fixed moment in time $t=constant$, the line element becomes 
\begin{equation}
ds^{2}=-\frac{dr^{2}}{1-\frac{b(r)}{r}}-(rd\varphi)^{2}.    
\end{equation}
Now, to represent the above equatorial planes as a surface embedded in a Euclidean space, we introduce the cylindrical coordinates as
\begin{equation}
 ds^{2}=-dz^{2}-dr^{2}-(rd\varphi)^{2},   
\end{equation}
which is straightforwardly rewritten as
\begin{equation}
ds^{2}=-\left[1+\left(\frac{dz}{dr}\right)^{2}\right]dr^{2}-(rd\varphi)^{2}.  
\end{equation}
Now, comparing the above equations, we have
\begin{equation}\label{eq31}
\frac{dz}{dr}=\pm \left[\frac{r}{b(r)}-1\right]^{-\frac{1}{2}},    
\end{equation}
where $z=z(r)$ represents the embedded surface. Intriguingly, when $r\rightarrow +\infty$ the above equation results in
\begin{equation}
{\frac{dz}{dr}~ \vline }_{+\infty}=0.    
\end{equation}

\begin{figure}[H]
    \centering
    \includegraphics[width=0.45\linewidth]{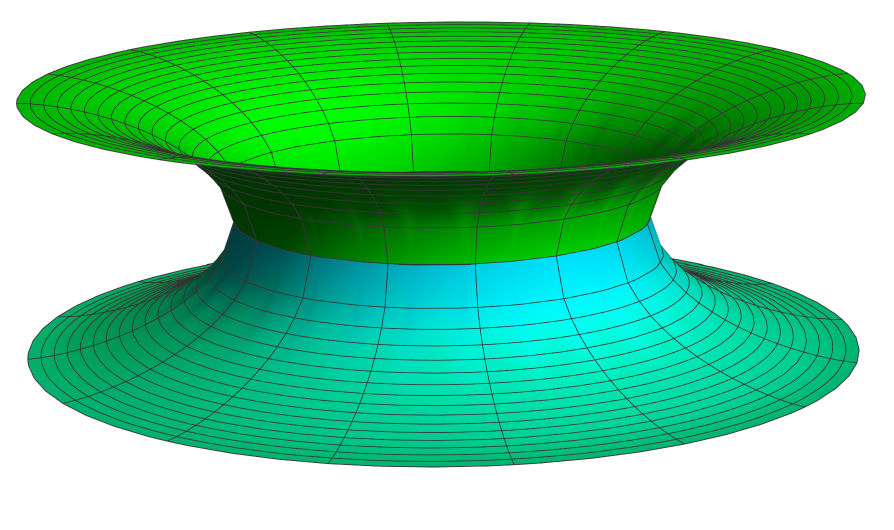}
    \includegraphics[width=0.45\linewidth]{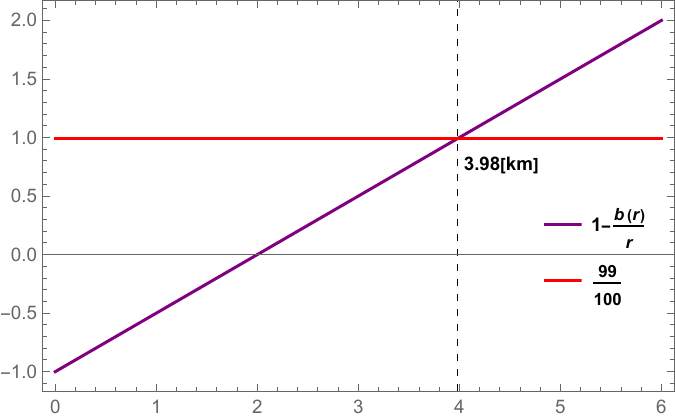}
    \caption{3D wormhole geometry using $z(r)$ function (left). The radial location $\hat{r} = |r_{\pm}|$ of spatial stations for the values $\{ r_{0}, \alpha \} = \{ 2, 2 \}$ (right).}
    \label{fig:placeholder}
\end{figure}
This shows that the embedding diagram contains two asymptotically flat patches. In order to illustrate the shape of the wormhole, a numerical treatment has been adopted. Based on the definition of a wormhole~\eqref{eq31} is divergent at $r=r_{0}$, indicating that the embedded surface is vertical there. In this case, we are assuming $\alpha=2$ for the embedded surface. On the left side of~[Fig.\ref{fig:placeholder}], the green colour represents the upper portion of the universe, while cyan represents the lower portion. A vertical axis is used to rotate the curve $z(r)$ in order to achieve the hyper-surface of the wormhole. Above the throat, the radial distance is positive, while below the throat, it is negative. Moreover, it is observed that the embedding surface becomes flat as one moves far away from the throat. 

According to Morris and Thorne \cite{Morris95},  the proper radial distance must act properly everywhere, i.e., $l(r)$ should be finite at all points in the space-time. This information can be obtained as the same method in \cite{Morris95}, which is that the factor $1-b(r)/r$ deviates from unity by no more than $1\%$. When this condition is implemented and the the numerical values $\{r_{0},\alpha\}=\{2,2\}$[km] are taken into account, one can obtain $\hat{r}=|r_{-}|=|r_{+}|\approx \pm 3.98$, so that,
\begin{equation}
l(r)=\pm \int_{r_{0}^{+}}^{\hat{r}}\left({1-\frac{b(r)}{r}}\right)^{-1/2}dr \approx \pm  3.97995
\end{equation}
On the right-hand side of~[Fig.\ref{fig:placeholder}], the radial location of the spatial stations can be observed.

\section{Conclusion}\label{Sec:V}
In this study, we investigated wormhole solutions within the framework of $f(R,\mathcal{L}_{m})$ gravity by considering the model $f(R,\mathcal{L}_{m}) = f_{1}(R) + \big[1+\lambda f_{2}(R)\big]\mathcal{L}_{m}$. The corresponding field equations were derived, and conditions were explored under which the density equation could yield a shape function consistent with the requirements for traversable wormholes. It was found that the energy density can take the form $\rho \sim 1/r^{2}$, and the associated shape function satisfies both the flare-out and traversability conditions. The obtained solution depends on the parameters $\alpha$ and $r_{0}$. For $\alpha=2$ and $r_{0}=2$, we derived explicit expressions for the energy density, radial pressure, and tangential pressure. The analysis of the energy conditions revealed that both radial and tangential components are satisfied within the parameter ranges $r \in [2,5]$ and $\lambda \in [2.6,4]$. As GR requires exotic matter to sustain traversable wormholes, in this modified gravity theory, we have demonstrated the existence of solutions that do not necessitate exotic matter at the wormhole throat. 

Finally, the corresponding wormhole geometry has been constructed, confirming the feasibility of traversable wormhole solutions within this modified gravity framework. A detailed analysis of the geometric properties has been carried out, with particular emphasis on ensuring that the proper radial distance $l(r)$ remains finite throughout the spacetime. This condition guarantees the internal consistency of the wormhole geometry and reinforces the physical viability of the obtained solutions.

\section*{Acknowledgement} 
ASA and RK would like to thank IUCAA for the opportunity to visit. BM acknowledges the support of Council of Scientific and Industrial Research (CSIR) for the project grant (No. 03/1493/23/EMR II).
\section*{Reference}
\bibliographystyle{utphys}

\bibliography{ref}
\end{document}